\documentclass[prl,aps, floatfix, superscriptaddress]{revtex4}
\usepackage{amsmath,bm,graphicx}
\usepackage{amssymb}
\usepackage{amsfonts}
\title{Geometrically-tuned channel permeability}

\begin{document}
\title{Geometrically-tuned channel permeability}
\author{Paolo Malgaretti}
\address{Max-Planck-Institut f\"{u}r Intelligente Systeme, Heisenbergstr. 3 D-70569 Stuttgart Germany \\ and IV. Institut f\"{u}r  Theoretische Physik, Universit\"{a}t Stuttgart, Pfaffenwaldring 57, D-70569 Stuttgart, Germany}

\author{Ignacio Pagonabarraga}
\address{Departament de Fisica Fonamental, Universitat de Barcelona Barcelona, C. Mart\'{i} i Franques 1, Barcelona, Spain}
\author{J.Miguel Rubi}
\address{Departament de Fisica Fonamental, Universitat de Barcelona Barcelona, C. Mart\'{i} i Franques 1, Barcelona, Spain}
\begin{abstract}
We characterize the motion of charged as well as neutral tracers, in an electrolyte embedded in a varying section channel. We exploit a set of systematic approximations that allows us to simplify the problem, yet capturing the essential of the interplay between the geometrical confinement provided by the corrugated channel walls and the electrolyte properties. 
Our simplified approach allows us to characterize the transport properties of corrugated channels when a net flux of tracers is obtained by keeping the extrema of the channel at different chemical potentials. For highly diluted tracer suspensions, we have characterized tracers currents and we have estimated the net electric current which occurs when both positively and negatively charged tracers are considered.  
\end{abstract}

\maketitle

\section*{I Introduction}

Both in biological situations and in synthetic devices charged particles are transported across a channel whose radial section ranges from the micro- to the nano-metric size. In such systems the confining walls are usually charged and the electrolyte organizes around the walls to screen their charges. As a result, an inhomogeneous electrostatic field develops inside the channels that  affects the transport  of charged tracers along the channel, leading to  a charge-dependent channel permeability. Until now, the control of such currents has generally relied on applying  an external driving whose origin is usually either of  electrical or hydrostatic origin. Tuning the external drive allows, for example,  for the control of particle currents as it happens in sodium-potassium pumping in neurons~\cite{albers}. 
However, the shape of the geometrical confinement  provides, in itself,  an alternative route to transport and current control. Different groups have shown that, in the presence of external forces, the local variation in channel section can induce novel dynamical regimes such as particle separation~\cite{Reguera2012,Motz,PaoloElecotrokinetics}, cooperative rectification~\cite{Malgaretti2012} and negative mobility~\cite{PaoloElecotrokinetics} just to mention a few among others~\cite{Bezrukov,Calero,StocastoResonance}. In particular, it has been shown that the inhomogeneous distribution of particles along the radial direction~\cite{Reguera2001} can modulate the overall current as it happens for neutral tracers under an external field, such as gravity.
Moreover, when an electrolyte is embedded in a varying-section channel, novel dynamical regimes, such as particle separation and negative mobility, can be attained when the Debye length (characterizing the decay of the electrostatic potential from the channel walls) and the channel section are comparable~\cite{PaoloElecotrokinetics}. Such a regime differs from previously studied scenarios~\cite{Hanggi,Ghosal} when the Debye length was much thinner than the channel's cross section.

Membrane permeability  depends  on the  membrane structure and also on how molecules diffuse  through it. It is in general not obvious how to disentangle to what extent the measured permeability is   essentially controlled by the  morphology or by the dynamics. 
Therefore, we will not consider the permeability associated to the flow through a pore in response to applied pressure gradients, as is usually analyzed in  the context of  porous materials. Rather, in  this piece of work, we will study the  permeability of a corrugated channel  driven by a difference in chemical potential at the ends of the channel. 
We will focus on the interplay between  diffusion and local  changes in the pore geometry, and will concentrate on the effect that channel shape has on the diffusion of ions inside the channel. In particular, we will not consider the permeability associated to the flow through a pore in response to applied pressure gradients, as is usually analyzed in  the context of  porous materials. Rather, in  this piece of work, we will study the  permeability of a corrugated channel  driven by a difference in chemical potential at the ends of the channel. Such a regime has the advantage that the fluid ions diffuse through can be considered as at rest therefore simplifying the overall dynamics. 

The permeability in ion channels, and in membranes, quantifies permeation across permeable substrates. In particular, the permeability coefficient  is an involved function that depends both  on intrinsic properties of the soluble molecules, such as their concentration and diffusion coefficient, on properties of the channel, such as its size and geometry, and in the interaction between the molecules and the channels, quantified for example through the partitioning coefficient (which quantifies the affinity of the molecule to the channel) or a slip coefficient. The Goldman-Hodgkin-Katz theory accounts for such ingredients and the  electrostatic interactions between molecules and charged ions and  predicts a strong, non-linear dependence of the permeability when a voltage difference is applied through the channel. This theory emphasizes the relevant contribution of the motion of the ions inside the channel to the channel permeability~\cite{ionchannels}. The results we will identify for the effective tracer diffusivity along the channel will provide  fundamental understanding on the impact that channel shape has on charged  ion permeabilities in both channels and membranes. Specifically, we shall show that the shape of the geometrical confinement provided by the channel walls as well as the electrolyte properties, captured by the Debye length, lead to a significant modulation of the overall channel permeability for both charged and neutral tracers. In particular, we find that the relative position of the bottlenecks as compared to the reservoirs at the end of the channel plays a key  role in the channel permeability, therefore providing additional ways to tune the net flux along the channel.    
The structure of the text is as follows: In Section II we will derive the Fick-Jacobs equation for charged tracers moving in a varying-section channel, in Section III we will present our results and finally in Section IV we will provide our conclusions.

\section*{II. Electrolyte embedded in a corrugated channel: a Fick-Jacobs approach}

We will consider   a symmetric, $z-z$ electrolyte embedded in a negatively charged channel of varying  half-section amplitude, $h(x)$, although analogous results  can be  obtained for an insulating channel, subject to a gradient of tracers'  concentration.  In order to describe the dynamics of suspended charged tracers, one needs to analyze the diffusion motion of the tracers and the electrostatic potential inside the channel, which constitutes a formidable task. However, the analysis  can be significantly simplified for highly diluted ionic concentrations and small $\zeta$ potential on the channel walls, i.e. $\beta\zeta e\ll 1$ where $\beta^{-1}=k_B T$ is the inverse temperature, with  $k_B$ the Boltzmann constant and $e$ the elementary charge. In such a regime, the Poisson-Boltzmann equation, which determines the electrostatic potential  in thermodynamic equilibrium,  can be linearized and  simplified  to the  Debye-Huckel equation for which the electrostatic potential decays exponentially with the distance from the wall over the characteristic   length scale, $\lambda\equiv k^{-1}$, namely the Debye length. We will also  consider  the regime where the charged solute suspension is highly dilute (tracer limit), when its dynamics is described by an advection-diffusion equation.  In order to gain insight in the properties of the channel permeability upon variation of the channel geometry, we will assume that the  channel half-section, $h(x)$, varies slowly, i.e. $\partial_x h(x) \ll 1$. Such an assumption allows for a  projection of the  convection-diffusion equation, which determines the dynamics of tracers inside the channel,  to an effective $1D$ equation, where the varying-section of the channel  enters as an entropic effective potential. This approximation, called Fick-Jacobs, has been used~\cite{Reguera2006,paolo_jcp_2013,StocastoResonance} and validated~\cite{Zwanzig,Reguera2001,Burada2007,Kalinay2008} in many different scenarios.

\begin{figure}
 \includegraphics[scale=0.5]{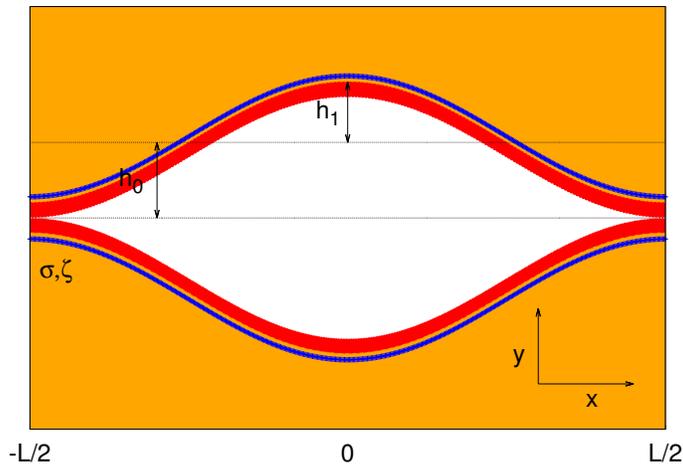}
 \caption{Electrostatic field inside a varying-section channel whose bottleneck half-amplitude $h_0-h_1$ is comparable with the Debye double layer thickness, $\lambda$.}
 \label{channel}
\end{figure}

We will consider a channel, of length $L$, whose section  varies only along the $x$-direction and it is constant along the $z$ coordinate, with a simple, periodic  shape whose half section is  determined by 
\begin{equation}
 h(x)=h_0+h_1\cos\left(\frac{2\pi x}{L}+\phi\right)
 \label{eq:profile}
\end{equation}
where $h_0$ and $h_1$ correspond to  the average and maximum channel modulation, respectively. In turn, $\phi$ determines the relative position of the channel bottleneck with respect to the connection of the channel to the reservoirs it is in contact with. For $\phi=0$, at $x=L/2$ the channel shows a bottleneck while for $\phi=\pi$ the minimal aperture of the channel occurs at $x=0$. 

The dynamics of tracers embedded in an electrolyte, quantified by its probability distribution inside the channel $P_{\alpha}(x,y,z,t)$,  with $\alpha =\pm,0$ denoting positive, negative and neutral tracers, is governed by  the convection-diffusion equation, which in the overdamped regime, reads 
\begin{equation}
\partial_{t}P_{\alpha}(x,y,z,t)=D\beta\nabla\cdot\left[P_{\alpha}(x,y,z,t)\nabla U_\alpha(x,y,z)\right]+D\nabla^{2}P_{\alpha}(x,y,z,t)
\label{smoluch}
\end{equation}
where $D$ is the tracer diffusion coefficient and $U_\alpha$ is the total   potential acting on the tracers,
\begin{eqnarray}
 U_\alpha(x,y,z)=\left\{\begin{array}{cc}
  q_\alpha e \psi(x,y,z), &  |y|\le h(x)\, \&\, |z| \le L_z/2 \\
 \infty, &  |y|> h(x)\, \mbox{or} \, |z|>L_z/2 \\
 \end{array}
 \right.
  \nonumber
 \label{potential}
\end{eqnarray}
that is periodic along the longitudinal direction, $U_\alpha(x,y,z)= U_\alpha(x+L,y,z)$, and confines the particles inside the channel. $\psi$ stands for the electrostatic potential inside the channel that, due to the high dilution of tracers is supposed to be unaffected by them, and $q_\alpha$ corresponds to the tracer valency while $e$ is the electron charge. We assume that all tracers have the same intrinsic diffusivity, $D$. Hence, the differences in tracer diffusion we will describe later will then be essentially due to the environment in which the tracers will move.
The electrostatic potential, $\psi$, obeys the Poisson equation
\begin{equation}
 \partial^2_x\psi(x,y,z)+ \partial^2_y\psi(x,y,z)+ \partial^2_z\psi(x,y,z)=-\frac{\rho_q(x,y,z)}{\epsilon}
 \label{poisson}
\end{equation}
where  $\epsilon$ is the liquid dielectric constant  and the value of the electrostatic potential at the interfaces depend on the  conducting nature of the channel walls. In the Poisson equation $\rho_q(x,y,z)$ stands for the local charge density of the electrolyte. At equilibrium, the charge density is derived using the Boltzmann distribution of ionic densities inside the channel. When the channel walls are smoothly-varying, $\partial_x h(x) \ll 1$, and assuming  lubrication $\partial^2_x\psi(x,y,z)\ll \partial^2_y\psi(x,y,z)$ and $\partial^2_z\psi(x,y,z)\ll \partial^2_y\psi(x,y,z)$, we can reduce the Poisson equation to an ordinary differential equation for the potential $\psi$ along the channel. At a solid, homogeneous\footnote{In the following we will consider either channel walls characterized by a constant potential, $\zeta$, or by a constant surface charge density, $\sigma$} interface, the electric field is perpendicular to the solid wall. As a result, for a varying-section channel the 
effective field along the channel corresponds to the  projection of the electric field along the wall. However, for a 
smoothly-varying channel amplitude, the projected  electrostatic field reads: $ E=E_0\cos(\theta)$, with $\theta=\arctan\left[\partial_x h(x)\right]$ the local  channel slope. Since, $\partial_x h(x) \ll 1$, the corrections on the electric field due to  changes in the channel  section are of second order in $\partial_x h(x)$ and can be safely neglected in the following.
For low salt concentrations and small $\zeta$ potential on channel walls, we can further simplify the Poisson equation by linearizing the charge density $\rho_q(x,y)\simeq \rho_0 \left(1-\beta ze \psi(x,y)\right)$, hence getting
\begin{equation}
 \psi(x,y,z)=\zeta\frac{ \cosh(k y)}{\cosh(k h(x))}
\end{equation}
for a channel made by conducting walls or 
\begin{equation}
 \psi(x,y,z)=\frac{\sigma}{2\epsilon}\frac{ \cosh(k y)}{\sinh(k h(x))}
\end{equation}
for an insulating channel characterized by a constant surface-density of electric charge $\sigma$. Such an assumption, known as Debye-H\"uckel approximation, allows to identify the screening length, $k^{-1}$, of the electrostatic potential as $k^{2}=\beta z e \rho_0/\epsilon$ where $\rho_0$ is the number density of the ions in the electrolyte and $z$ their valency. The approximation made for the electrostatic field reflects in the Debye length, $\lambda$, that results  constant up to second order  in $\partial_x h(x)$. Finally, for $\partial_x h(x) \ll 1$ we can approximate the transverse profile of the probability distribution function (pdf), $P_{\alpha}(x,y,z,t)$, of a tracer of valency $q_\alpha$ by its profile at equilibrium, i.e., we can factorize the pdf by assuming~\footnote{The prefactor in Eq.~\ref{free-en} has been introduced to keep the argument of the logarithm dimensionless and its value does not affect neither the probability profile nor the probability flux (see Appendix)}: 
\begin{eqnarray}
P_{\alpha}(x,y,z,t) & = & p_{\alpha}(x,t)\frac{e^{-\beta q_\alpha e \psi(x,y)}}{e^{-\beta A_\alpha(x)}} \label{eq:p-alpha}\\
\beta A_\alpha(x) & = & -\ln\left[\frac{1}{2h_0L_z} \int_{-L_z/2}^{L_z/2}\int_{-h(x)}^{h(x)}e^{-\beta q_\alpha e \psi(x,y)}dy dz\right]. 
\label{free-en}
\end{eqnarray}
After integrating over the channel cross section we arrive at 
\begin{equation}
 \dot p_{\alpha}(x,t)=\partial_x D\left[ \beta p_{\alpha}(x,t)\frac{\partial A_\alpha(x)}{\partial x}+\partial_x p_{\alpha}(x,t)\right].
\label{FJ1}
\end{equation}
This  expression encodes both the confining as well as the electrostatic potential in the effective potential $A(x)$ whose shape, see Eq.~\ref{free-en}, resembles that of an equilibrium free energy.
Since all the quantities of interest are independent of $z$, without loss of generality we can assume $\int_{-L_z/2}^{L_z/2}dz=1$ and consider all quantities per unit of transverse length, $L_z$. Defining the average, $x$-dependent, electrostatic energy as
\begin{equation}
\langle V_\alpha(x)\rangle=e^{\beta A_\alpha(x)}\int_{-h(x)}^{h(x)}q_\alpha e \psi(x,y)e^{-\beta q_\alpha  e \psi(x,y)}dy
\label{avg-V}
\end{equation}
from the definition of $A_\alpha(x)$ we can define the, dimensionless, entropy along the channel as $k_BTS_\alpha(x)=\langle V_\alpha(x)\rangle -A_\alpha(x)$, from which we can define 
\begin{equation}
 S_\alpha(x)=\ln\left[\frac{1}{2h_0}\int_{-h(x)}^{h(x)}e^{-\beta q_\alpha e \psi(x,y)}dy \right]+\beta\left\langle V_\alpha(x)\right\rangle.
\end{equation}
In the linear regime, $\beta q_\alpha e \psi(x,y) \ll 1$, we can linearize the last expression getting
\begin{equation}
 S(x)\simeq \ln\left(\frac{2h(x)}{2h_0}\right),
\end{equation}
where the entropy has a clear geometric interpretation, being the logarithm of the space, $2h(x)$, accessible to the center of mass of a point-like tracer. Accordingly,  we introduce the entropy barrier, $\Delta S$, defined as
\begin{equation}
\Delta S=\ln \left(\frac{h_{max}}{h_{min}}\right),
\label{entropy-barrier}
\end{equation} 
which represents the difference in the entropy  evaluated at the maximum, $h_{max}$, and minimum, $h_{min}$, of the channel aperture.

\section*{III Tracer dynamics}

A net,  constant  flux of tracers, of magnitude $J$,  characterizes the steady state  motion of tracers in the channel whenever there is a difference in the  chemical potential  of the  two baths the channel is in contact with.  Since there is no electrostatic potential difference between the reservoirs at the two channel ends, and the  tracer density is small, the tracer chemical potential in the reservoirs reads $\mu_\alpha = k_BT \ln \rho_\alpha$. If $\mu_\alpha$ differs at the ends of the channel, the associated $\Delta \mu_\alpha$ will lead to  a tracer flux along the channel. We are interested in understanding the dependence of channel permeability on the channel geometry as a result of the imposed chemical potential difference.

From Eq.~(\ref{FJ1}), in steady state we can express the tracer density  per unit length  profile inside the channel, $n_\alpha(x)$, as
\begin{equation}
n_{\alpha}(x)=e^{-\beta A_\alpha(x)}\left[-\frac{J_{\alpha}}{D}\int_{-\frac{L}{2}}^{x}e^{\beta A_\alpha(z)}dz+\Pi_{\alpha}\right],
\label{eq:density}
\end{equation}
where we have used that  for the tracers $p_{\alpha}(x)$ and $n_{\alpha}(x)$ are proportional to each other~\footnote{We choose the tracer density, $n_\alpha(x)$, instead of the tracer probability, $p_{\alpha}(x)$, because it provides a more direct connection with the chemical potential in the  reservoirs the channel is in contact with. The density in such reservoirs constitutes the natural control parameter}.  From $n_\alpha(x)$, the local  average tracer density can be finally obtained  as $\rho_{\alpha}(x)=\frac{n_{\alpha}(x)}{2h(x)}$; this relation provides the  natural  link with the tracer densities in the reservoirs. Specifically, we will consider that the reservoirs at the two ends of the channel are kept at tracer densities $\rho_{\alpha,1}$ and $\rho_{\alpha,2}$, corresponding to an imposed chemical potential difference $\Delta \mu_{\alpha} = \ln \frac{\rho_{\alpha,1}}{\rho_{\alpha,2}}$. Imposing $n_{\alpha}\left(-\frac{L}{2}\right)=2\rho_{\alpha,1}h(-\frac{L}{2})$ and $n_{\alpha}\left(\frac{L}{2}\right)=2\rho_{\alpha,2}h(\frac{L}{2})$ we determine the  
two constants in Eq.~(\ref{eq:density}))
\begin{equation}
\Pi_{\alpha}=2h\left(L/2\right)\rho_{\alpha,1}e^{\beta A_\alpha\left(-\frac{L}{2}\right)}
\end{equation}
\begin{equation}
J_{\alpha}=-2Dh\left(L/2\right)\frac{(\rho_{\alpha,2}-\rho_{\alpha,1})}{L}\frac{e^{\beta A_\alpha\left(\frac{L}{2}\right)}}{\frac{1}{L}\int_{-\frac{L}{2}}^{\frac{L}{2}}e^{\beta A_\alpha(x)}dx}
\label{eq:flux}
\end{equation}
where in the last expression we have exploited the symmetry of the channel, $h(-\frac{L}{2})=h(\frac{L}{2})$ that, using Eq.(~\ref{free-en}), implies $A_\alpha(-\frac{L}{2})=A_\alpha(\frac{L}{2})$. Eq.~\ref{eq:flux} provides the dependence of the tracers' mass flow   due to both the gradient in tracer concentration, $\nabla \rho_{\alpha} \equiv (\rho_{\alpha,2}-\rho_{\alpha,1})/L$, and the geometry of the channel. 
We can rewrite Eq.~\ref{eq:flux} as:
\begin{equation}
 J_{\alpha}=-2Dh\left(L/2\right)\nabla\rho_\alpha\chi_\alpha
 \label{eq:flux-1}
\end{equation}
where the geometrical dependence on the flux is encoded in the dimensionless parameter
\begin{equation}
 \chi_\alpha=\frac{e^{\beta A_\alpha\left(\frac{L}{2}\right)}}{\frac{1}{L}\int_{-\frac{L}{2}}^{\frac{L}{2}}e^{\beta A_\alpha(x)}dx}
 \label{eq:chi}
\end{equation}
According to Eqs.~\ref{eq:flux-1},\ref{eq:chi}, $\chi_\alpha$ captures the dependence of tracers' flux upon both channel geometry and tracers' charge. Moreover, Eq.~\ref{eq:chi} shows that $\chi_\alpha$ is independent on the magnitude of the drive, $\nabla\rho_\alpha$, responsible for the onset of the flux. In particular, when $\chi_\alpha>1$, channel corrugation enhances tracers' flow as compared to the case of a flat channel hence leading to a  larger channel permeability while a reduced permeability will be obtained for $\chi_\alpha<1$.

Eq.~(\ref{eq:chi}) shows that for constant channel sections $\chi_\alpha=1$ irrespectively to tracers' charge. Therefore, according to Eq.~\ref{eq:flux-1} when the channel section is constant tracers' flux, and hence channel permeability, are insensitive to tracer charge, provided they undergo the same chemical potential gradient, $\nabla\rho_\alpha$. Tracer charge will determine the   partitioning of the tracer inside the channel and its distribution in the channel section, but will not affect its effective diffusivity.

The dependence of $\chi_\alpha$ on the channel varying geometry  is quite involved, as can be appreciated in Eq.~\ref{eq:chi}. We can gain insight into the impact that channel corrugation has on its permeability by analyzing the particular case where the  channel amplitude varies   linearly, namely $h(x)=h_0-h_1|x|/L$. In this geometry,   $\beta A(x)\simeq -\ln(h_0/L)+ (h_1/h_0)|x|/L$. This regime can be achieved, for example, if the channel corrugation is weak and the electrostatic potential varies linearly. Or, more generically, one can consider a channel whose section does not change strongly and hence one can identify the coupled effect of channel corrugation and electrostatic potential variation through an effective, uniform variation along the channel. 

For neutral tracers this regime  corresponds, exactly, to a channel with  a linearly increasing cross section. For charged tracers, in the regime of  a very narrow double layer, $kh_0\gg 1$, tracers are essentially uniformly distributed in the channel section; accordingly,  they will behave as neutral tracers.  Under these assumptions we have
\begin{equation}
\int_{-\frac{L}{2}}^{\frac{L}{2}}e^{\beta A(x)}dx=2\int_{0}^{\frac{L}{2}}e^{-\ln h_0/L+h_1/h_0x}dx=2L^2\frac{e^{\beta h_1/(2 h_0)}-1}{h_1}.
\end{equation}
Substituting the last expression in Eq.~\ref{eq:chi} and using the fact that $A\left(\frac{L}{2}\right)=A\left(-\frac{L}{2}\right)=\ln h_{0}+h_1/(2 h_0)$ we get
\begin{equation}
\chi_\alpha=\frac{1}{1-e^{-\beta h_1/(2 h_0)}}
\end{equation}
Finally, recalling that in this case $\beta\Delta A=h_1/(2h_0)$ the last expression reads
\begin{equation}
\chi_\alpha=\frac{1}{1-e^{-\beta\Delta A}}
\label{eq:flux_simpl}
\end{equation}
Interestingly the last expression shows that the permeability does not depend  symmetrically on $\Delta A$. Hence, the permeability of a concave channel, where the minimal aperture is in the center of the channel, will be smaller than the complementary channel where the center of the channel has the largest aperture. For a channel with the same mean section, the difference in the permeability for channels with opposed  curvature will differ exponentially with the channel corrugation.

It is insightful to analyze  the  limits
\begin{eqnarray}
\lim_{\Delta A\rightarrow0}\chi_\alpha & = & 1\\
\lim_{\Delta A\rightarrow\infty}\chi_\alpha & = & \beta\Delta A\\
\lim_{\Delta A\rightarrow-\infty}\chi_\alpha & = & 0,
\end{eqnarray}
which clearly show the strong dependence of the permeability on the effective, free energy barrier $\Delta A$. This free energy barrier, which encodes the change in channel section and the local modifications this induces in the  electrostatic potential, controls the impact  of the environment on the net transport properties of charged tracers.

\begin{figure}
\includegraphics[scale=0.45]{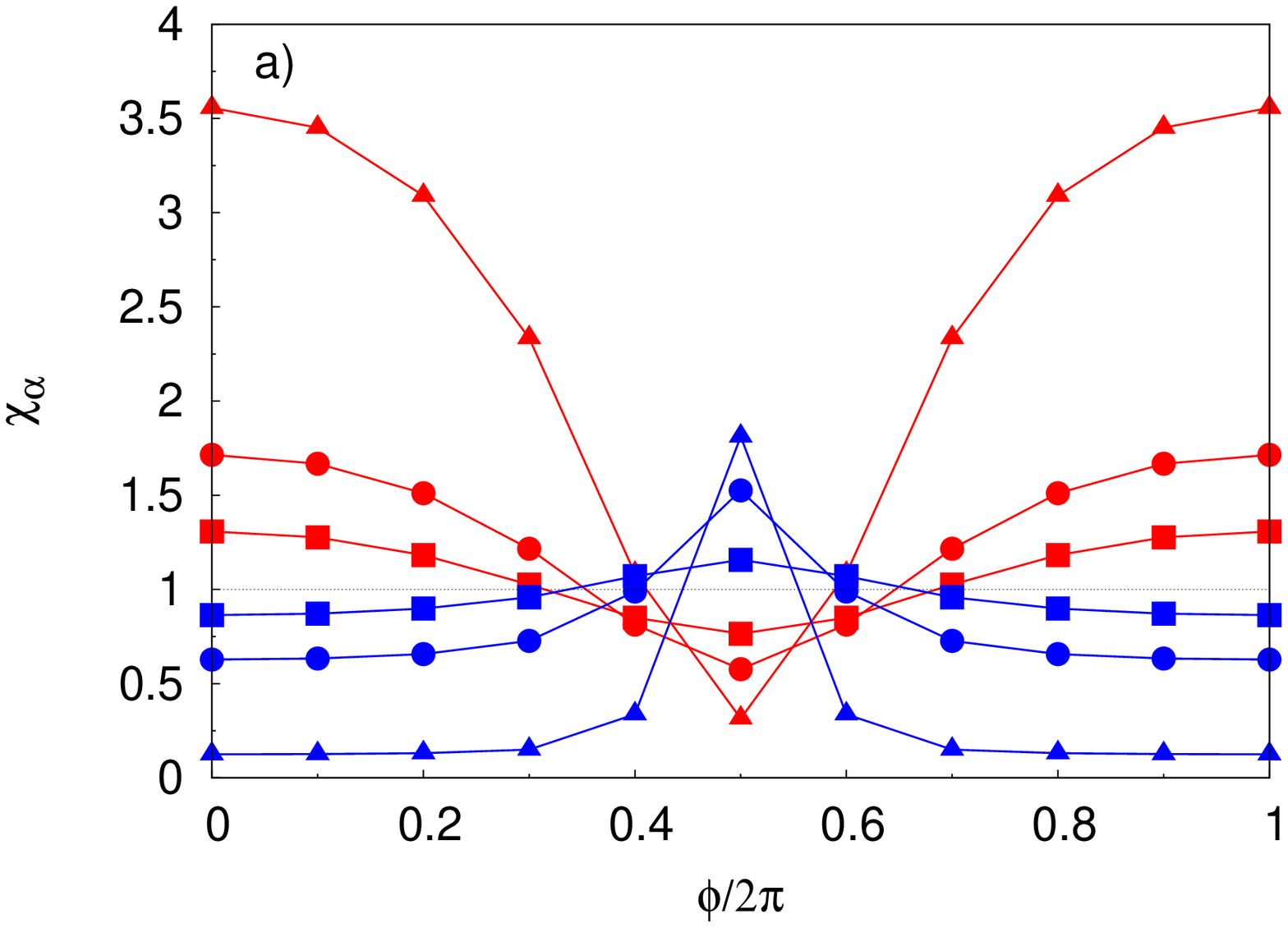} \includegraphics[scale=0.45]{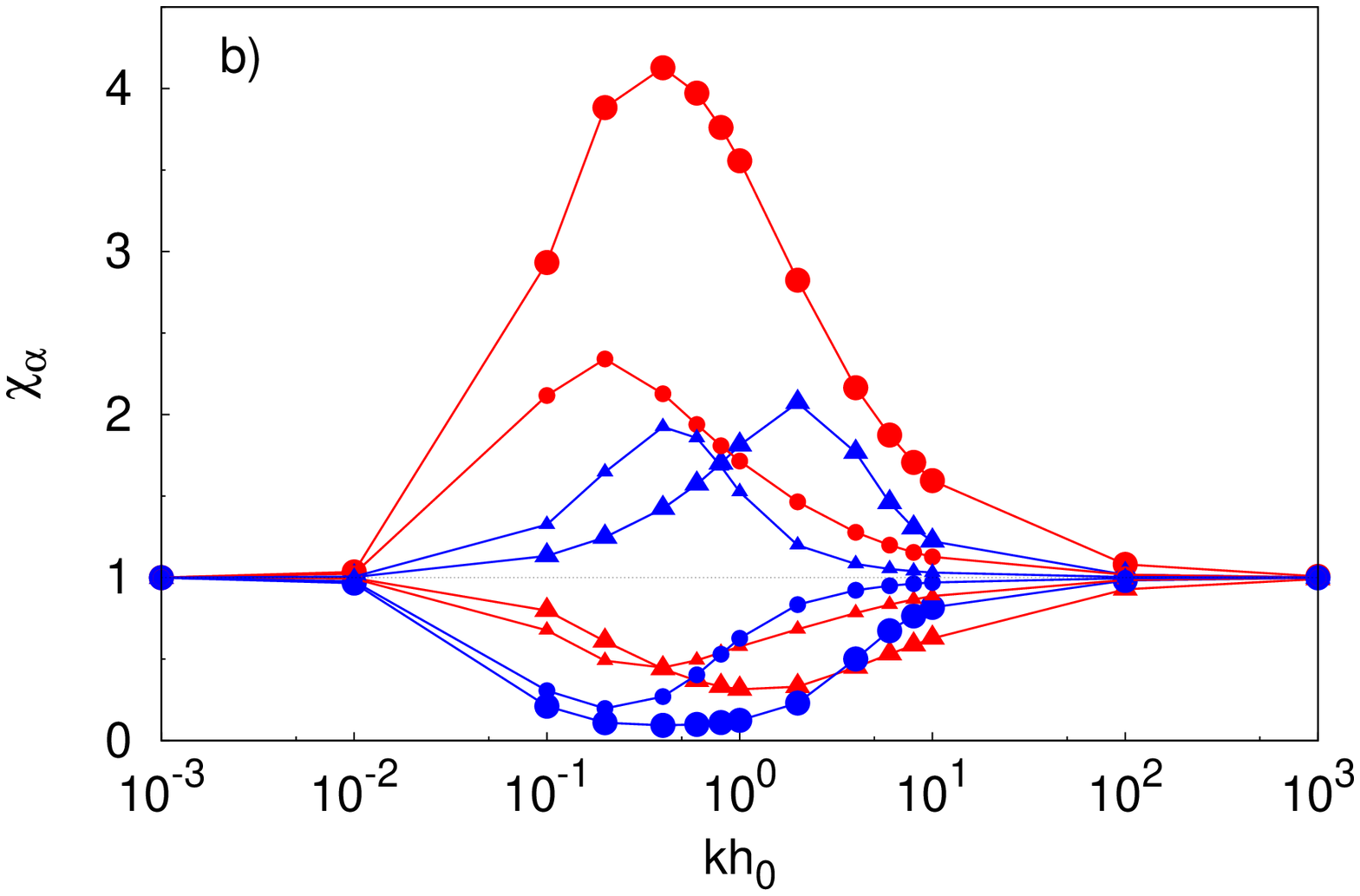}
\includegraphics[scale=0.45]{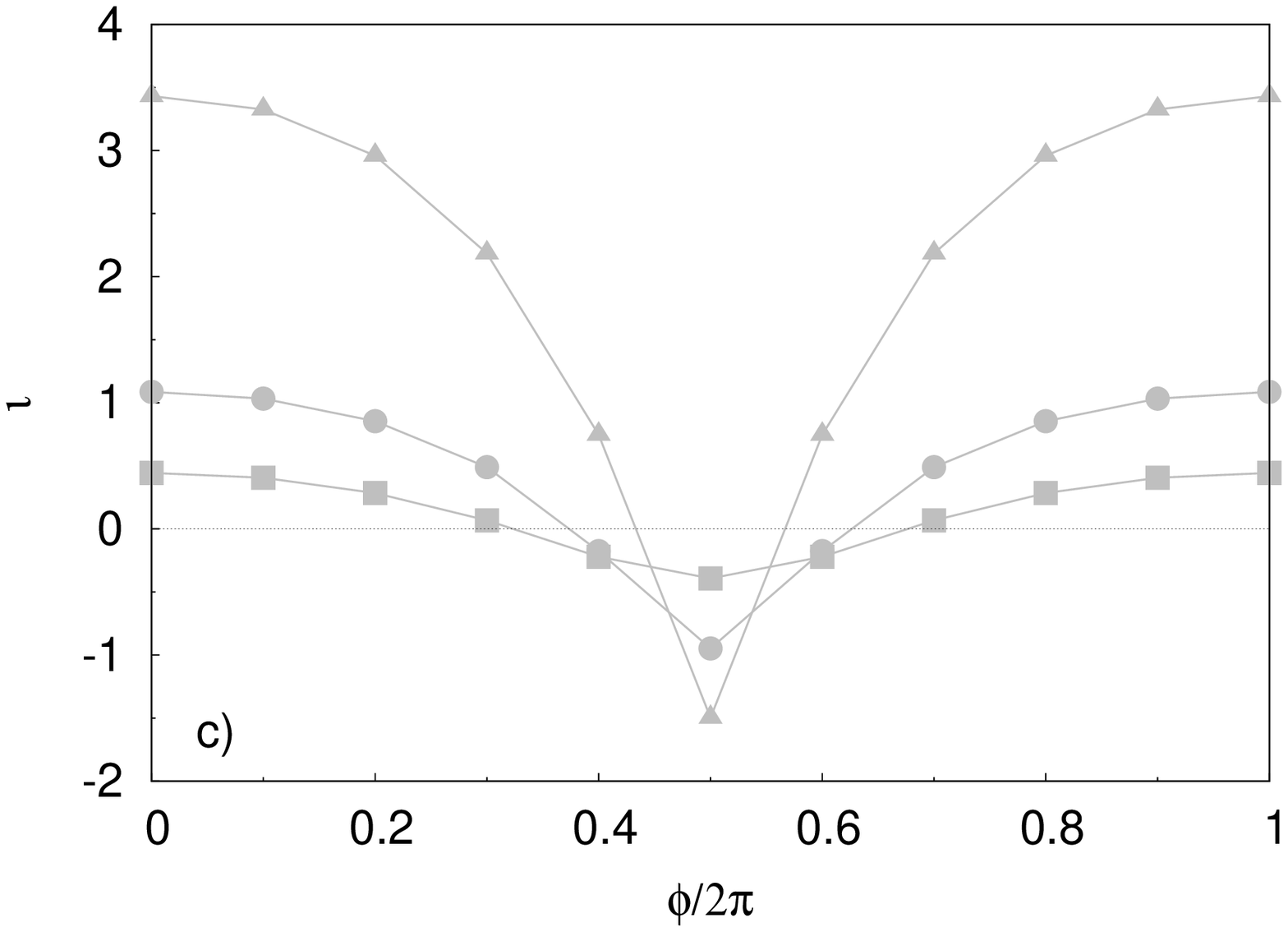} \includegraphics[scale=0.45]{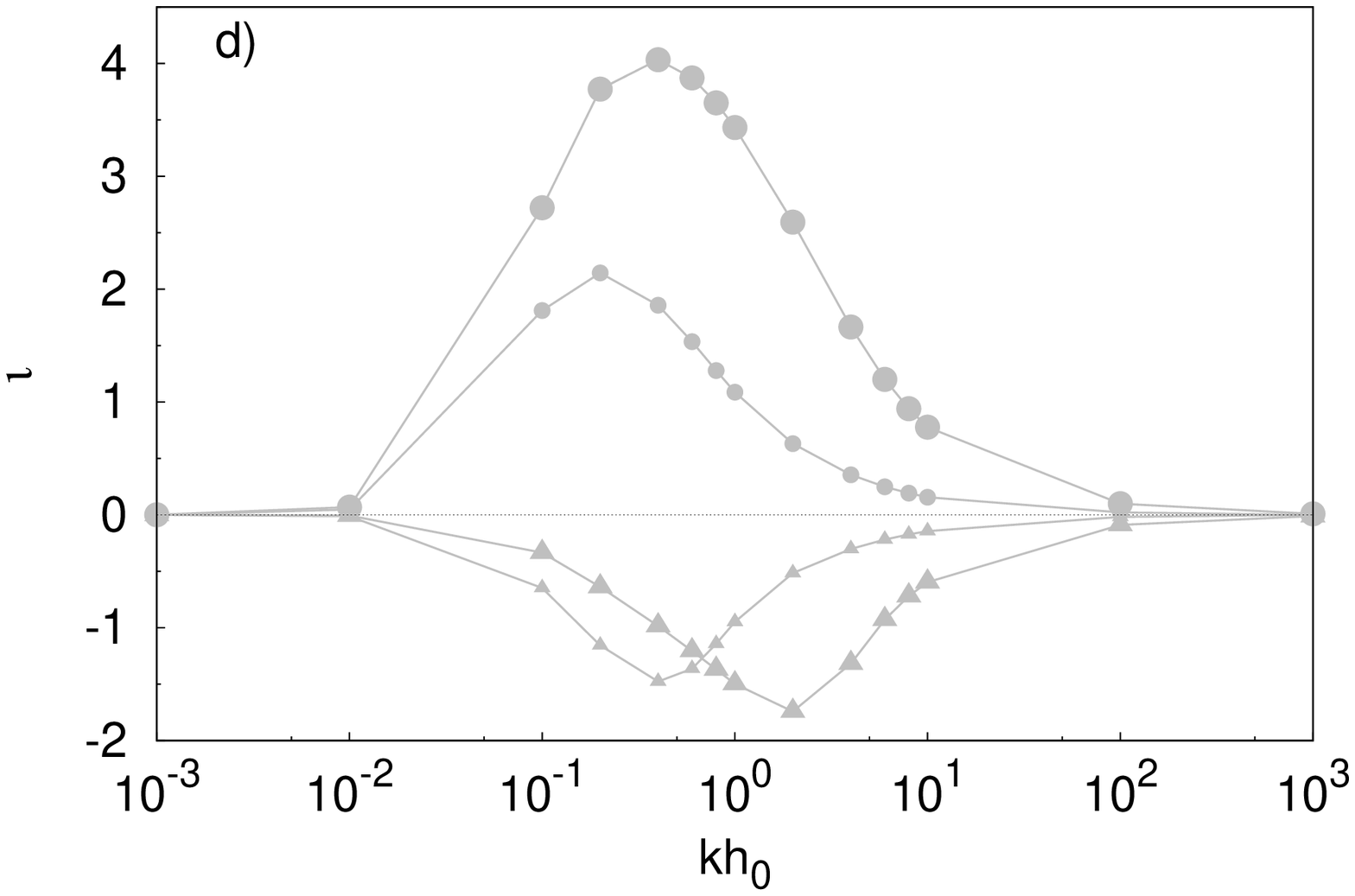}
\caption{Tracers permeability. a: $\chi_\alpha$ as a function of the phase $\phi$ for positive (red) and negative (blue) tracers, normalized by the current of neutral tracers,  for different values of the corrugation, $\Delta S=1.1,2,3.7$ for squares circles and triangles respectively with $kh_0=10$. b: $\chi_\alpha$ for charged tracers, normalized by the permeability of neutral tracers $\chi_0$, as a function of the Debye length, $\lambda$, expressed in units of $kh_{0}$ for different channel geometries: $\phi=0$ ($\phi=0.5$) is characterized by circles (triangles) while the color code captures the charge of the tracers red (blue) stand for positive (negative) tracers while marker size is proportional to the amplitude of the corrugation $\Delta S=2,3.7$. c-d: electric current, as defined in Eq.~\ref{eq:def-curr}, calculated from the data of the corresponding above panels.}
\label{fig:current}.
\end{figure}

\section*{IV Results}
In order to characterize quantitatively the dependence of the channel diffusivity as a function of tracer charge and  for different channel geometries and electrolyte concentrations, we have numerically evaluated Eq.~\ref{eq:chi}. For the simple channel profile, characterized by Eq.~(\ref{eq:profile}), neutral tracer diffusion does not depend on channel shape. Hence, the deviations we  will analyze emerge from a correlation between channel corrugation and the local variations in electrostatic potential. As  shown in Fig.~\ref{fig:current}.a, $\chi_\alpha$ depends on channel shape and the relative position of the maximum aperture with respect to the channel ends, quantified by the  phase shift $\phi$.  The variation indicates that $\chi_\alpha$, and hence the permeability,  depends symmetrically  on the distance of the  channel maximum with respect to the channel ends. The impact of channel shape on tracers' dynamics has 
opposite trends depending on tracers' charge. Negative tracers experience a net repulsion form the solid surfaces and will preferentially accumulate in the channel center. They display a maximum in permeability when the bottleneck is at channel edges, while the minimum in the permeability happens in the opposite situation. Positive tracers exhibit a complementary behavior. These tracers accumulate preferentially at the channel walls and  benefit from the shape when the charge accumulated at the two walls do not interfere with each other. The non-monotonous behavior of the permeabilities leads to different scenarios where the ratio between the permeabilities of charged and neutral tracers can vary significantly.  We can identify different channel configurations for which one of the three species has the largest permeability. For example, positive tracers (hence attracted from the negatively charged walls) experience the maximum permeability when the channel shows a bottleneck at $x=0$, while negatively 
charged tracers are fastest in the case of bottlenecks at channel edges. Alternatively, for $\phi\simeq 0.4,0.6$ neutral tracers are faster than both positively and negatively charged tracers.

Charged tracer permeabilities are strongly affected by the electrolyte properties, which are captured by the Debye length, $\lambda$, and depend on the combined properties of the channel and the electrolyte, as it has been shown for the case of electrostatically driven tracers~\cite{PaoloElecotrokinetics}.  Fig.~\ref{fig:current}.b shows that if $\lambda$ and the channel average amplitude, $h_0$, are not comparable, the modulation in the permeability due to the geometrical confinement vanishes and both positively and negatively charged tracers experience the same permeability. For $k h_0 \ll 1$ the electrolyte is uniform in the channel section, while for  $k h_0 \gg 1$ the electrolyte is  localized at the wall and the  fluid inside the channel is essentially  neutral. In both cases all tracers behave analogously and become insensitive to channel corrugation. In contrast, when $kh_0\simeq 1$ the permeabilities shows maxima (minima) according to tracers' charge. The regime $kh_0\simeq 1$ can be 
obtained in nanofluidic devices for which the Debye length, that is typically $\lambda\simeq 1-10\, nm$) matches the typical amplitude, $h_0$ of the channel. Alternatively the same regime can be achieved in microfluidic devices by exploiting low polar solvents characterized by larger Debye lengths, e.g. $\lambda\simeq 1.6 \mu m$ for CHB-decalin~\cite{Leunissen2007} Accordingly, Fig.~\ref{fig:current}.b  shows that negatively charged tracers experience larger permeabilities for $\phi=0.5$ than for $\phi=0$ (compare the blue big (small) triangles and blue big (small) circles) while the opposite holds for positively charged tracers that experience a larger permeability when the bottleneck is at the middle of the channel ($\phi=0$). The qualitatively different response of positively and negatively charged tracers to channel corrugation indicates the relevance of the geometrical  coupling together with the electrostatic attraction or repulsion to the walls; such a dependence is encoded in the free energy 
dependence, $\Delta A$. 
The analysis at the end of the previous section  illustrates how a change in the  sensitivity to spatial variations of the effective free energy has a strong impact in tracer diffusivity.  The results obtained highlight the relevance of the regime where the Debye length and channel section are comparable in size.

Since  tracers move with different effective  diffusivities along the channel, depending on their charge, tracer motion as a result of the applied density gradient will induce also  a net electric current. This is an electric current induced by the tracers because the electrolyte is in equilibrium, and corresponds to a coupled transport effect induced only by geometrical  variations~\footnote{We assume that electroneutrality is achieved at both ends of the channel yet we impose a density contrast at channel ends for both positively and negatively charged tracers.}. For simplicity, we will consider that the channel is subject to the same density gradient for negative, positive and neural tracers, $\Delta \mu_+=\Delta \mu_-=\Delta \mu_0$. In this case, rather than  computing the electric current itself, $I=J_+-J_-$, it is insightful to   analyze the tracer electric current relative to the reference mass current of neutral tracers,
\begin{equation}
 i=\frac{I}{J_0}=\frac{\chi_+-\chi_-}{\chi_0},
 \label{eq:def-curr}
\end{equation}
which, according to Eq.~(\ref{eq:chi}), can be expressed in terms of the corresponding effective tracer permeabilities.
Fig.~\ref{fig:current}.c shows the relative  electric current as a function of the position of the maximum aperture of the channel. The results sow that a relevant current can be induced by the tracer density gradient and that the current can change its  sign depending on the channel geometry. The  channel where  the relative electrostatic current is positive correlates with the situations where the diffusion of positive tracers is enhanced, as could be expected. Therefore, it it is possible to control the magnitude and the sign of the electric current by properly tuning the position of the channel bottleneck.  In particular, the maxima of the electric current are obtained for $\phi=0,\phi=\pi$, i.e. when the bottleneck is at the center of the channel or at channel edges and the absolute value of the current is maximized in the bottleneck case ($\phi=0$). Hence,  the shape of the channel controls the amplitude of the net electric current induced by the tracers density contrast imposed at channel ends. Fig.~\
ref{fig:current}.d shows that the amplitude of the current is modulated by the Debye length, $\lambda$, and the maximum is achieved when is comparable to the characteristic  channel section. Fig.~\ref{fig:current}.d also shows that this current vanishes when the double layer is much thinner or wider than the channel aperture. In this situation the  inhomogeneous electrolyte distribution in the channel is negligible. This feature highlights the fact that the induced tracer current emerges from the combined geometric modulation and the charge distribution across it.
Therefore, varying the channel geometry and the electrolyte ionic strength it is possible to control both he magnitude and the sign of the electric current

\begin{figure}
\includegraphics[scale=0.45]{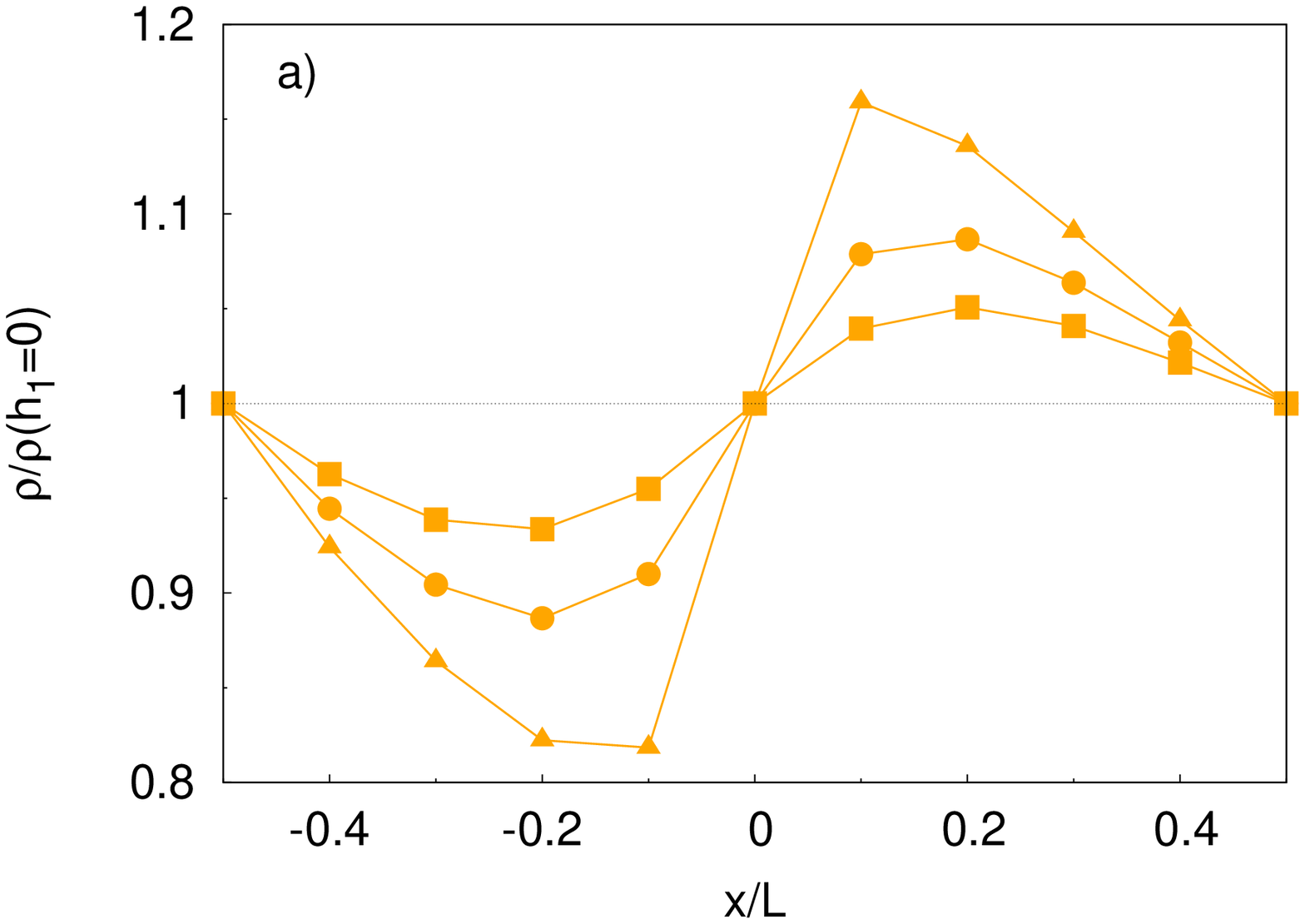} \includegraphics[scale=0.45]{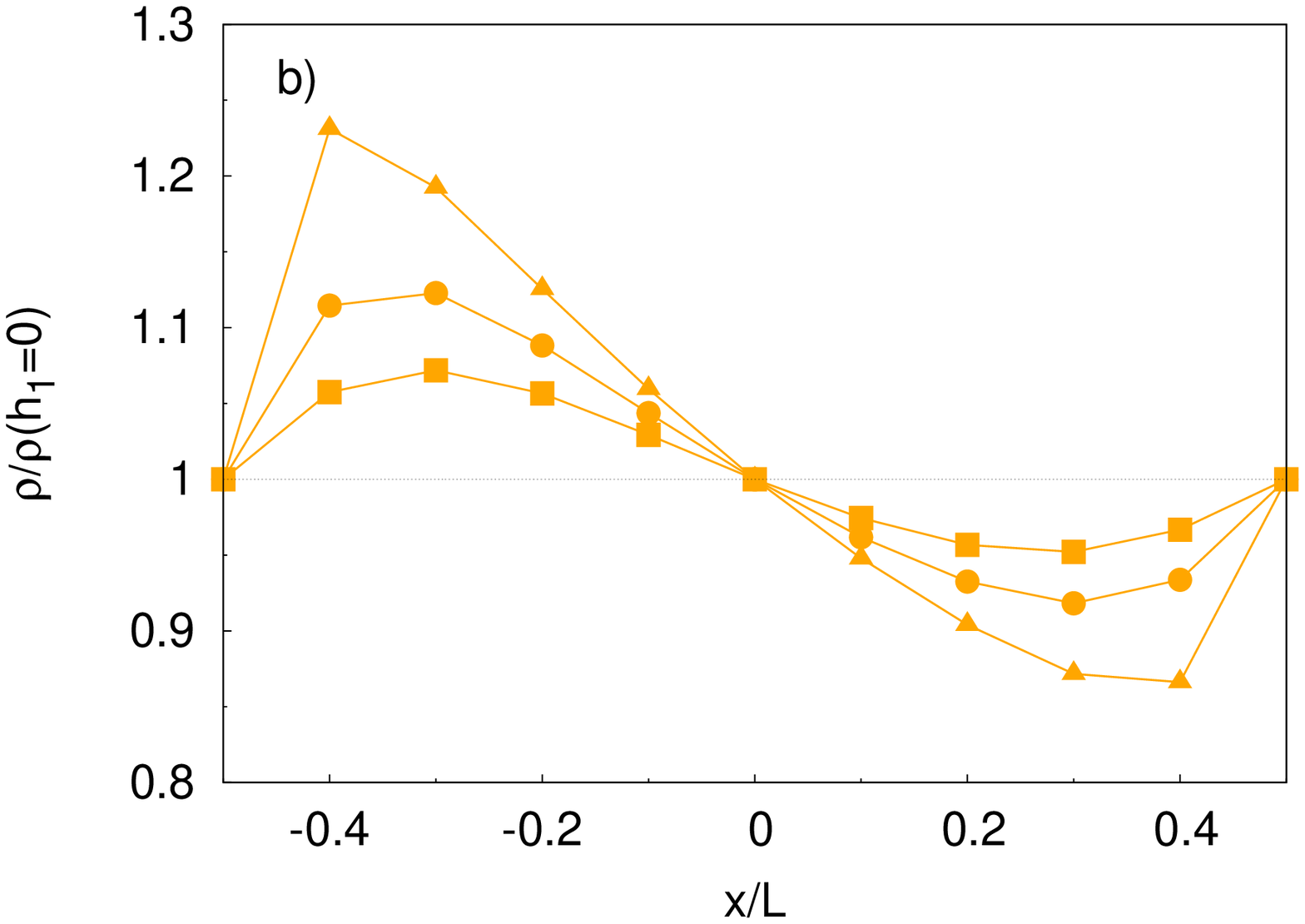}
\caption{Neutral tracers density profile, normalized by the density profile in the case of a flat channel $\rho(h_1=0)$, for different channel geometries $\Delta S=1.1,2,3.7$ for squares, circles and triangles respectively characterized by, $\phi=0$ (left), $\phi=0.5$ (right).}
\label{fig:null}
\end{figure}

The modulation in channel geometry as well as the magnitude of the  Debye length affects also the tracer distribution along the channel. Fig.~\ref{fig:null} shows that even for neutral tracers, for which the density profile is independent of $\lambda$, the density profile is affected by the geometrical constraint. In particular, in the case of a flat channel, the  chemical potentials contrast imposed at the ends of the channel leads to a linear density profile inside the channel (data not shown). On the contrary, when the section of the channel varies, we observe a departure from the linear profile that is asymmetric with respect to channel geometry. In particular, for $\phi=0$ ($\phi=0.5$) we observe an excess of accumulation (depletion) of particles close to the reservoir characterized by the larger chemical potential while the opposite holds in the vicinity of other reservoir. Interestingly, the excess of particle density is sensitive to the derivative of channel section; 
the maximum of the deviations from the linear profile is obtained when $\partial_x h(x)$ is maximum. 

\begin{figure}
\includegraphics[scale=0.45]{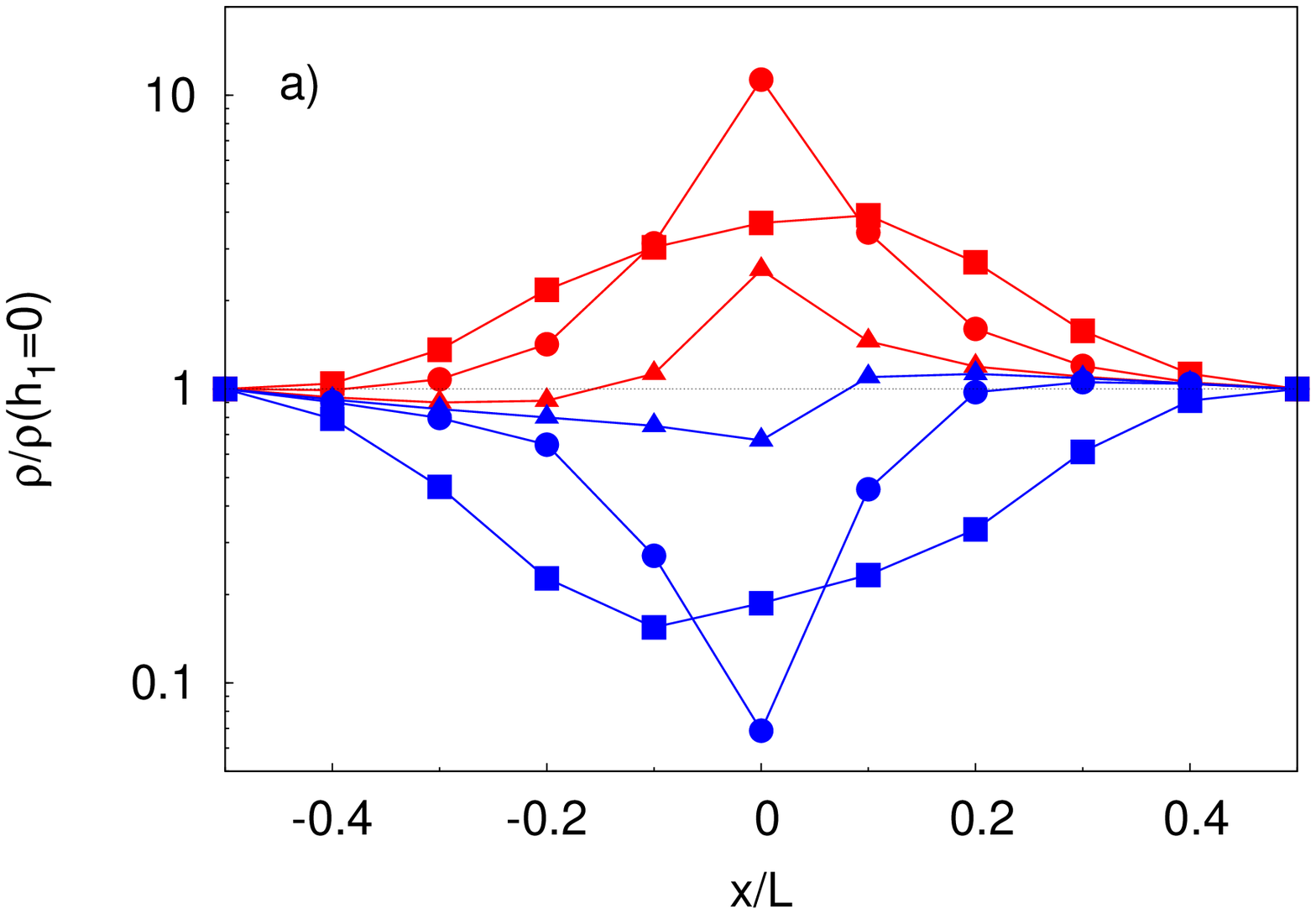} \includegraphics[scale=0.45]{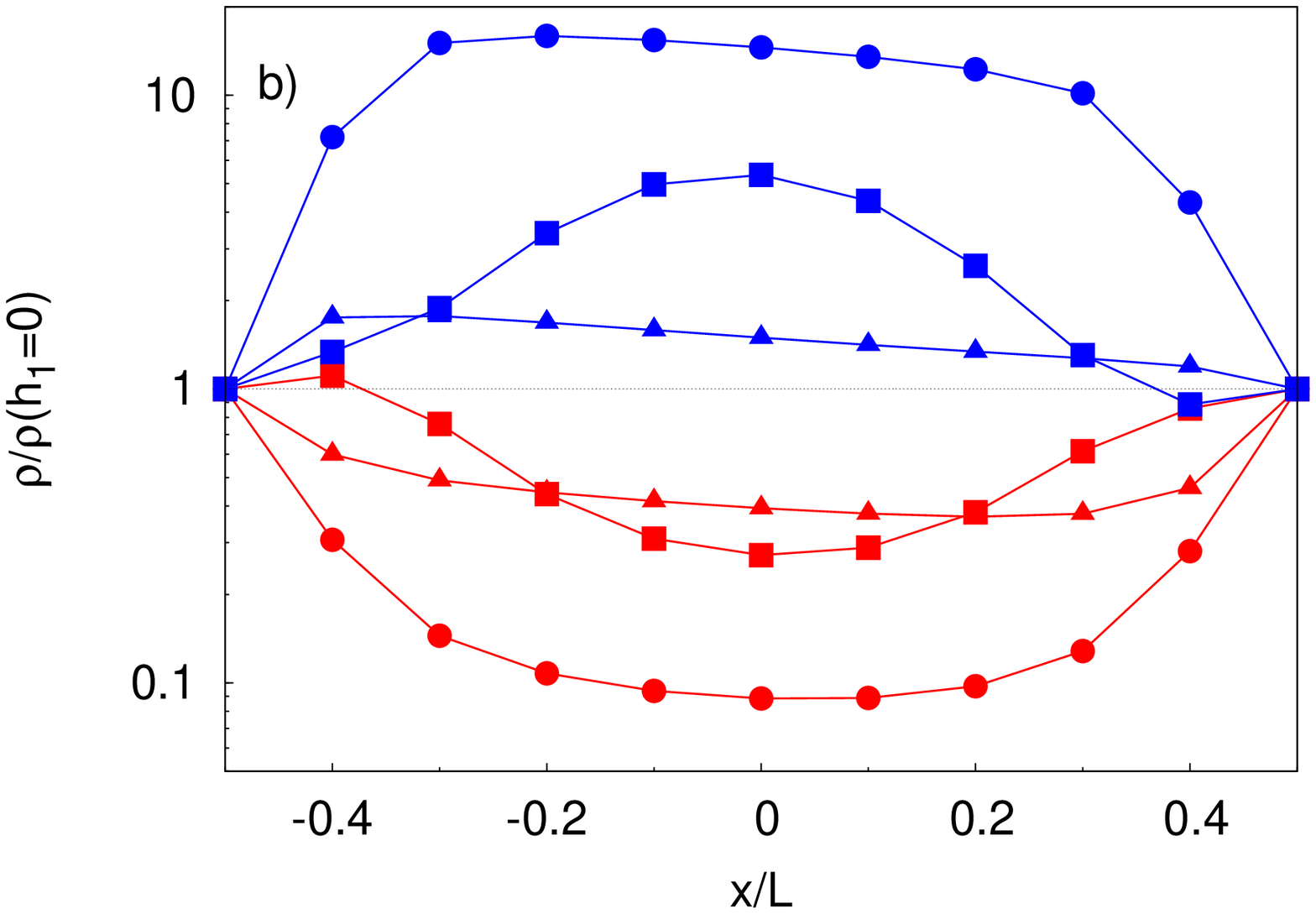}
\caption{Charged tracers density profile, normalized by the density profile in the case of a flat channel $\rho(h_1=0)$, for positive (red markers) and negative (blue markers) tracers and different channel geometries $\Delta S=1.1,2,3.7$ for squares, circles and triangles respectively characterized by, $\phi=0$ (left), $\phi=0.5$ (right).} 
\label{fig:charge} 
\end{figure}

When tracers are charged, the effect of the confinement on tracers' density is amplified. Fig.\ref{fig:charge} shows an enhanced accumulation (depletion) as compared to the case of neutral tracers. In particular, positively charged tracers, (attracted by the negatively charged channel walls) tend to accumulate for $\phi=0$, i.e. when the channel bottleneck is at $x=0$ while the opposite holds for negatively charged tracers. The amplitude of the excess accumulation/depletion of tracers has a non-trivial dependence on channel geometry. In particular, comparing the two panels of Fig.\ref{fig:charge} we see that the density profiles for the two cases, namely $\phi=0$ and $\phi=0.5$ have different shapes and the amplitude of the modulation in the density profile is larger for $\phi=0.5$. 
The electrolyte properties play a relevant role in determining tracer density profiles. As shown in Fig.~\ref{fig:charge}, the maximum amplitude of the modulation  in the density shows a non-monotonous behavior with $\lambda$ and its maximum is obtained for $kh_{0}\simeq 10$, i.e. when the amplitude of the bottlenecks, $k(h_0-h_1)\simeq 0.5$  is comparable to the Debye length.

\section*{V Conclusions}

We have studied the  impact of the channel shape on the permeability of charged and neutral tracers.  We have identified a novel mechanism that controls tracer permeability  due to variations in the channel geometry. 
In particular, we have studied the geometrical dependence of the channel permeability for neutral and charged tracers, when the system is driven by means of a chemical potential contrast imposed by controlling tracers concentration at channel ends. In order to keep analytical insight, we have assumed that the channel amplitude varies smoothly, $\partial_x h(x)$, along its longitudinal axis, $x$. In such a regime, a  lubrication approximation is reliable and allows for a significative simplification of the analysis. Moreover, in such a regime it is possible to exploit the Fick-Jacobs approximation that allows us to factorize the distribution probability of tracers, hence reducing the problem to a $1D$ problem whose solution can be accessed analytically. 

We have exploited such a framework to characterize the permeability of a corrugated channel for the transport of positive and negative charged and neutral tracers. For these systems, we have found that the relative magnitude of the permeabilities of both positively and negatively charged tracers as well as neutral tracers can be tuned by the geometry of the channel, leading to a variety of scenarios where positively (negatively) charged tracers, can experience larger or smaller permeabilities than neutral ones. 
As it has already been shown in the case of electric driving forces~\cite{PaoloElecotrokinetics}, the geometrically-induced control on tracers  current is sensitive to electrolyte properties captured by the Debye length, $\lambda$. The  sensitivity of tracer diffusion to channel corrugation is maximum when $\lambda$ is  comparable to the characteristic channel section.  Such a 
regime $kh_0\simeq 1$ can be obtained in nanofluidic devices for which the Debye length, that is typically between $1$ and $10\, nm$, matches the typical amplitude, $h_0$ of the channel, or in microfluidic devices by exploiting low polar solvents characterized by larger Debye lengths such as $\lambda\simeq 1.6 \mu m$~\cite{Leunissen2007}. In contrast, for very wide and narrow double layers, i.e. for $kh_0 \gg1$ or $kh_0\ll1$,  the impact of channel corrugation vanishes asymptotically. The interplay between channel corrugation and electrolyte structuring across it leads to a new cross transport mechanism. Specifically, we have found that when the Debye length is comparable to the channel section, a net electric current develops when a chemical potential gradient of positive and negatively charged tracers acts at the channel ends. This corrugation-induced cross transport effect can be of significance in the electric transport through ionic channels and deserves to be analyzed in detail. 

\begin{acknowledgments}
J.M. R. and I.P. acknowledge  the Direcci\'on General de Investigaci\'on (Spain) and DURSI project for financial support
under projects  FIS\ 2011-22603 and 2014SGR-922, respectively. J.M. R.  and I.P. acknowledges financial support from {\sl Generalitat de Catalunya } under program {\sl Icrea Academia}. P.M. acknowledges Dr. Adam Law for carefully reading the manuscript.
\end{acknowledgments}
\appendix*

\section*{Appendix A}
In order to keep the argument of the logarithm in Eq.~\ref{free-en} dimensionless, we have introduced a prefactor, $1/2h_0L_z$. In the following we show that the choice of this prefactor is arbitrary since it will not affect either the probability distribution $p_\alpha$ or its flux.  The starting point are Eqs.~\ref{smoluch},\ref{eq:p-alpha},\ref{free-en} that we report here:
\begin{equation}
\partial_{t}P_{\alpha}(x,y,z,t)=D\beta\nabla\cdot\left(P_{\alpha}(x,y,z,t)\nabla U_{\alpha}(x,y,z)\right)+D\nabla^{2}P_{\alpha}(x,y,z,t)
\end{equation}
with the ansatz: 
\begin{eqnarray}
P_{\alpha}(x,y,z,t) & = & p_{\alpha}(x,t)\frac{e^{-\beta q_{\alpha}e\psi(x,y,z)}}{e^{-\beta A_{\alpha}(x)}}\\
\beta A_{\alpha}(x) & = & -\ln\left[\frac{1}{2h_{0}L_{z}}\int_{-L_{z}/2}^{L_{z}/2}\int_{-h(x)}^{h(x)}e^{-\beta q_{\alpha}e\psi(x,y,z)}dydz\right].
\end{eqnarray}
We can rewrite the last equation as $\beta A_{\alpha}(x)=-\ln\left[\frac{1}{2h_{0}L_{z}}\right]-\ln\left[\int_{-L_{z}/2}^{L_{z}/2}\int_{-h(x)}^{h(x)}e^{-\beta q_{\alpha}e\psi(x,y,z)}dydz\right].$
From the last expression we can calculate $e^{-\beta A_{\alpha}(x)}=e^{-\beta A_{0}}e^{-\beta A_{1}(x)}$
where $A_{0}=\ln\left[\frac{1}{2h_{0}L_{z}}\right]$and $A_{1}(x)=\ln\left[\int_{-L_{z}/2}^{L_{z}/2}\int_{-h(x)}^{h(x)}e^{-\beta q_{\alpha}e\psi(x,y)}dydz\right]$.
Substituting the last expressions in Eq.1 and integrating in $y$ we get:
\begin{equation}
\dot{p}\int dy\frac{e^{-\beta q_{\alpha}e\psi(x,y)}}{e^{-\beta A_{0}}e^{-\beta A_{1}(x)}}=D\beta\partial_{x}\int_{-\infty}^{\infty}p_{\alpha}(x,t)\frac{e^{-\beta q_{\alpha}e\psi(x,y)}}{e^{-\beta A_{0}}e^{-\beta A_{1}(x)}}\partial_{x}U+D\partial_{x}^{2}p_{\alpha}(x,t)\frac{e^{-\beta q_{\alpha}e\psi(x,y)}}{e^{-\beta A_{0}}e^{-\beta A_{1}(x)}}dy
\end{equation}
where in the last step we have neglected the terms like $\partial_{y}\int[...]dy$
since they provide null contributions. From the last equation it is
clear that we can get rid of the term $e^{-\beta A_{0}}$ since, being
independent on both, $x$ and $y$, it can be taken out of the integrals
and differentiation operators. Recalling that $\int dy\frac{e^{-\beta q_{\alpha}e\psi(x,y)}}{e^{-\beta A_{1}(x)}}=1$ and 
performing the integral in the last expression we get: 
\begin{equation}
\dot{p}_{\alpha}(x,t)=\partial_{x}D\left[\beta p_{\alpha}(x,t)\frac{\partial A_{1}(x)}{\partial x}+\partial_{x}p_{\alpha}(x,t)\right]
\end{equation}
Since $\partial_{x}A_{1}(x)=\partial_{x}A_{\alpha}(x)$ we can rewrite
the last expression as 
\begin{equation}
\dot{p}_{\alpha}(x,t)=\partial_{x}D\left[\beta p_{\alpha}(x,t)\frac{\partial A_{\alpha}(x)}{\partial x}+\partial_{x}p_{\alpha}(x,t)\right]
\end{equation}
showing that the normalization does not affect the evolution of the tracer probability along the channel. 

\bibliography{letter_electrokinetics}

\end{document}